# Room temperature mid-IR single photon spectral imaging

Jeppe Seidelin Dam, Christian Pedersen and Peter Tidemand-Lichtenberg

*DTU Fotonik, Technical University of Denmark, DK-4000 Roskilde, Denmark*

**Abstract**

Spectral imaging and detection of mid-infrared (mid-IR) wavelengths are emerging as an enabling technology of great technical and scientific interest; primarily because important chemical compounds display unique and strong mid-IR spectral fingerprints revealing valuable chemical information. While modern Quantum cascade lasers have evolved as ideal coherent mid-IR excitation sources, simple, low noise, room temperature detectors and imaging systems still lag behind. We address this need presenting a novel, field-deployable, upconversion system for sensitive, 2-D, mid-IR spectral imaging. Measured room temperature dark noise is 0.2 photons/spatial element/second, which is a billion times below the dark noise level of cryogenically cooled InSb cameras. Single photon imaging and up to 200 x 100 spatial elements resolution is obtained reaching record high continuous wave quantum efficiency of about 20 % for polarized incoherent light at 3 $\mu$m. The proposed method is relevant for existing and new mid-IR applications like gas analysis and medical diagnostics.

Optical spectroscopy within the UV, Visible and NIR has for decades been an indispensable method for identification and quantization of chemical analytes. However, emerging mid-IR applications like environmental gas monitoring or in life science call for improved detection systems challenging today's capabilities in terms of sensitivity and/or imaging functionality. A few examples; In face of global warming, mid-IR detectors capable of measuring minute gas concentrations are needed since important green house gasses like carbon dioxide ($CO_2$), Carbon monoxide (CO), methane ($CH_4$), and nitrous oxide ($N_2O$) have their fundamental absorption bands located in the mid-IR[1]. CO as an example requires a detection sensitivity in the order of about 100 ppb[2]. Monitoring atmospheric trace molecules at these levels provides important input to climate models used for studying global warming and its consequences for life on earth[3]. In life science, the spectral regime from 0.3-2 µm has already been utilized for fundamental studies of breath analysis. However significant improvements can be expected from mid-IR spectroscopy[4,5]. On-line detection of the numerous different molecules (> 1000) existing in the exhaled human breath may lead to new non-invasive diagnostics tool for doctors. However, such biomarkers are frequently below ppb (parts/billion) in exhaled breath. Exhaled concentration of Ethane (at 3.3 µm) for instance is investigated as a marker for asthma and chronic obstructive pulmonary disease present in the 100 ppt (parts/trillion) level pinpointing the necessity for highly sensitive methods[5]. Likewise 1-butanol and 3-hydroxy-2-butanone in breath could be useful breath biomarkers for lung cancer[6]. In the 3-15 µm wavelength regions, 2-D mid-IR spectral imaging demonstrates potential for identifying cancerous tissue providing a new tool for cancer diagnostics. In this wavelength region each organic compound and functional group has a well characterised spectral fingerprint of vibronic rather than electronic nature, making identification and quantification easier[7,8].

Frequency upconversion of infrared images was investigated already in the 1960's and 1970's[9,10,11,12,13,14,15,16,17]. However, the research field was practically abandoned 30 years ago due to prohibitively low conversion efficiencies. The quantum efficiency (QE) to our knowledge never exceeded $2 \times 10^{-7}$ for continuous wave (cw) systems[13]. Recently[18], we demonstrated a quantum efficiency of $2 \times 10^{-4}$ for incoherent image upconversion with 200x1000 pixel elements from red to the blue spectral region. These results demonstrated the potential of the technology[19]. Cw upconversion in recent years[20,21,22,23] has not been directly used for infrared imaging, however waveguide single spatial mode upconverters has been used for infrared photon detection[24]. In upconversion spectroscopy low noise devices in the near infrared has been reported[25,26,27,28].

Today, the preferred method for infrared spectroscopy is Fourier Transform Infrared spectroscopy (FTIR). However, the complexity and cost associated with FTIR can be prohibitive for many applications. Mid-IR cameras are presently based either on micro-bolometer technology or InSb semiconductor technology with low energy band gap. Even at cryogenic temperatures, these IR cameras suffer from large dark noise[29].

In this work, we demonstrate for the first time a field deployable rugged image upconversion device, which can be attached directly to a regular CCD camera. The upconversion device has a measured quantum efficiency (QE) of 20 %. This is 6 orders of magnitude improvement over the early cw IR image upconversion experiments. The system can be tuned to any wavelength between 2.85 and ~5 µm, and easily acquires images of thermal light sources, even at live video frame rates. We report an experimentally determined dark noise at room temperature of only 0.2 photons/spatial element/second, which combined with a near infrared (NIR) camera with single-photon detectivity, provides the first mid-IR single photon imaging device to date. A novel method providing monochromatic images from the upconversion process is outlined, pointing towards high-resolution multispectral imaging.

To demonstrate the spectroscopic functionality, we present upconversion images of light emission from a candle at specific wavelengths in the 2.9 to 4.2 μm range, showing the spatial distribution of soot, hydrocarbons (C-H), water vapor ($H_2O$), and carbon dioxide ($CO_2$), respectively.

The device is shown in Fig. 1. The wavelength conversion device can be tuned to any wavelength in the 2.85 to 5 μm region, converting the infrared image to the ~800 nm wavelength range for subsequent simple low-noise detection using a silicon based camera. The considered wavelength interval is dictated by periodically poled Lithium Niobate (pp-LN) which is transparent up to about 5 μm. The lower spectral limit is set by the available poling periods for our non-linear crystal.

Our approach consists of different elements as discussed in the following. A diode-pumped solid state laser is designed to generate an intense, continuous, $TEM_{00}$, intra-cavity laser beam. In the present configuration, this is achieved using a $Nd:YVO_4$ laser cavity oscillating at 1064 nm, pumped by a 4 W, 808 nm broad area laser diode (BAL). See Fig. 1c. A 20 mm long, 1 mm thick pp-LN crystal is placed inside the laser cavity. The pp-LN is phase-matched for Sum Frequency Generation (SFG) between the 1064 nm laser field and the wavelengths of the incoming mid-IR radiation. Including the intra-cavity non-linear crystal, a circulating power of up to 100 W is generated. The QE scales linearly with mixing power, hence, the high circulating power leads to an increase in QE by roughly a factor of 50-100 over a corresponding single-pass configuration. The QE of this set-up is calculated to be 20 % for correctly polarized, on-axis and phase-matched light[30]. The expected QE is confirmed by experiments as described in the methods section below and constitute to our knowledge state-of-the-art in cw, image upconversion efficiency. The improved QE is a corner stone in the practical exploitation of upconversion technology. The improvement in quantum efficiency since the early work is primarily due to the use of highly transparent periodically poled non-linear crystals and a low loss intra-cavity design.

The upconverted image formation can be thought of as imaging through an small soft aperture. However, in addition, the light changes it's wavelength. The size of the soft aperture determines the obtainable image resolution. The transverse dimension of the soft aperture is given by the beam diameter of the 1064 nm mixing laser, hence, the Fourier transform of the laser field gives the point spread function. Thus, larger transverse dimensions of the mixing laser will result in higher spatial resolution[30].

In the present system an image resolution of ~200 x 100 resolvable image elements is experimentally obtained from upconversion of thermal light at 3 μm, transmitted through a standard resolution target. The spatial resolution is lower in the vertical direction of the image, since the aperture in vertical direction is limited to avoid total internal reflections from top and bottom of the ~10 mm wide by 1 mm thick non-linear crystal. These reflections can be avoided if necessary by bonding an index matched transparent material to the pp-LN's upper and lower surface. This approach will also allow use of a larger laser beam diameter, resulting in improved image resolution. The image conversion takes place in an infinity corrected plane, thus image content is carried as angular information. The advantage is that no image degradation caused by the finite length of the non-linear crystal is necessary[14, 30]. This choice of geometry does, however, result in different propagation angles inside the crystal phase-match at different wavelengths. This in consequence leads to encoded spectral information in the acquired images. We will later investigate a method to exploit or circumvent this feature.

To demonstrate the image conversion capability, images at various mid-IR wavelengths from sample combustion processes (butane flame or a candle) were recorded. See Fig. 2. These images can easily be acquired at live video frame rates. At around 2.9 µm, water vapor exhibits significant emission at flame temperature. Since the spectrum of water vapor contains several nearly equidistantly spaced emission lines, a unique ring pattern is generated, where each ring in the pattern corresponds to light emission of a specific wavelength. Thus, the image of the butane flame will appear brighter in regions associated with the emission lines of water vapor. This illustrates imaging with spectral information, as seen in Fig. 2a. The image intensity distribution matches well with the theoretically calculated spectrum of hot water vapor, see Fig. 2b. The phase-matched wavelength depends on the propagation angle inside the crystal as shown in Fig. 2c. Figure 2d shows an image of a candle where thermal emission from soot (the yellow part of a candle flame) and the wick is also clearly seen. The presence and location of these rings is a unique marker of hot water vapor. As seen in the video, moving the candle leaves the position of the ring pattern stationary as expected. Figure 2e shows an image where the system is tuned to ~3.4 µm where hydrocarbons have a strong emission band. This allows us to directly image hot vaporized hydrocarbons in the candle flame – which, as expected, appear in a small region around the wick. Likewise, Figure 2f shows an image at a wavelength where $CO_2$ has strong emission bands (~4.2 µm), thus indicating the existence of hot $CO_2$ in the entire flame. Hot $CO_2$ can be traced far above the flame (30 cm or more).

Thus, by imaging at different wavelengths, the chemistry of a candle flame can be identified. We emphasize, that while the chemistry of the candle is well-known, it is a convenient test sample for demonstrating the spectral properties of the imaging system. The bandwidth of the upconversion system varies as the phase matched wavelength is changed. At 2.9 µm the bandwidth is 5 nm, gradually increasing to 10 nm at 3.4 µm, and 25 nm at 3.8 µm, while the bandwidth increases sharply to 200 nm at 4.2 µm, since the first order term in the phase-match condition vanishes there. Above 4.2 µm the bandwidth decreases again.

As discussed above, a specific phase-match condition, determined by the non-linear crystal parameters, leads to a phase-match curve as shown in Fig. 2.c. This implies, when using an infinity corrected geometry, that a given pixel element in the image plane is transformed into a unique angle in the crystal plane, which, according to Fig. 2.c, will allow a hereto unique wavelength to be efficiently upconverted. Thus by varying the phase-match condition of the non-linear crystal, corresponding to a vertical translation of the phase-match curve depicted in Fig. 2.c, we can by image reconstruction from these individually acquired images, piece together a set of images, each containing light from only one specific narrow band of wavelengths. This novel idea leads to spectral decomposition of a multi-spectral image or even hyper-spectral imaging. We demonstrate the image reconstruction technique in a simple case using monochromatic illumination. In Fig. 3 we illuminate a cross test target with a narrowband 3 µm laser. By changing the temperature of the non-linear crystal, we can upconvert different sections of the cross. The obtained frames can then be added together to obtain a full, reconstructed image as shown in Fig. 3d and the accompanying video. The efficiency of the conversion process remains the same for each image.

If fast tuning is required, the phase-match condition can be varied by other means than temperature tuning Other means include rotation of the non-linear crystal, electro optic tuning, or tuning of the mixing laser wavelength.

Dark noise is inherent to IR detectors relying on an absorption process since any finite temperature of the absorbing detector material leads to thermal radiation which is then reabsorbed and converted into noise electrons. Dark noise at room temperature is especially pronounced in the mid-IR and IR range as evident from the Planck blackbody radiation curve. The dark noise of a cryogenically cooled InSb mid-IR camera is typically in the order of a hundred million electrons per pixel per second[31]. In contrast to this, detectors for visible light can have practically zero dark noise since the band gaps for visible detectors are well above the energy levels of interfering room temperature thermal emission. The method presented here, is based on image upconversion. Since the upconversion material is ideally transparent[32] in the IR wavelength region of interest, it will not emit thermal radiation[33] even at room temperature. The low noise upconverted images are conveniently located around 800 nm where Si-based cameras are highly effective. Combining image upconversion and modern EM-CCD cameras, mid-IR single photon images can be obtained. In Fig. 4 theoretical estimates, based on blackbody radiation, and the experimentally obtained dark noise values are compared. The emissivity of the crystal[32] at 3 µm is 0.5 %/cm. The actually measured dark noise follows the theoretical predictions well, and we conclude that the dark noise in the present system is approximately 0.2 photons/spatial element/second, when operating at 30 °C. In our specific set-up, this corresponds to 0.008 electrons/pixel/second. The dark noise of the camera itself (Andor Luca S) is specified to 0.05 electrons/pixel/second. In order not to be limited by the about 1 electron read noise we tested a high end, low noise, camera. This camera (Andor iXon3 897) is able to measure single photons, which allowed us to confirm the expected low upconversion noise levels, and make the, to our knowledge, first single photon sensitive mid-infrared camera. We emphasize that upconversion was achieved at room temperature, whereas the camera chip was cooled to -85°C. The noise levels are expected to decrease even further if the upconversion device is also cooled. Since the noise depends on Planck radiation at the phase-matched wavelength and the transparency of the crystal, generally the noise levels are expected to increase for longer wavelengths.

A major result in this work is a greatly improved quantum efficiency reaching 20 % for on-axis, phase-matched optimally polarized light. The low QE previously demonstrated for upconversion imaging has until now prohibited this technology for real applications. The quantum efficiency could be increased beyond 50 % using a more powerful pump laser, a higher finesse laser cavity, or more efficient/longer non-linear crystals.

In the present work, we achieved 200 x 100 resolvable image elements. The resolution in our set-up is limited by the aperture size defined by the laser beam diameter. The laser beam diameter is limited by the thickness by which the pp-LN crystals can be produced. Using an astigmatic laser beam, the image resolution on one axis can be improved[18]. A second method involves bulk crystals. Here we are not technically limited by the limitations in poling but only by the physical crystal sizes, and will furthermore be able to extend the wavelength range deeper into the infrared. However, this will trade off versatility given by periodic poling in the design process.

In conclusion, we have constructed a compact, field deployable system for mid-IR imaging, compatible with standard visible and near infrared cameras. We have shown spectral imaging in the mid-IR by capturing spectrally resolved images at selected wavelengths in the mid-IR range. We point to several applications of the technology including astronomy, life sciences, gas imaging, industrial vision and security. By spectral decomposition we have outlined and demonstrated a method for extracting quasi-monochromatic image information leading towards hyper-spectral imaging. We have demonstrated superior noise characteristics without cryogenic cooling. At 3 μm the dark noise added by the upconversion process was a mere 0.2 photons/spatial element/second, thus comparable to high quality imaging devices for visible light. This demonstration of single photon infrared imaging enables new low light level mid-IR applications, using room temperature turn-key devices.

**Methods**

**Experimental system.**

The imaging wavelength conversion module consists of a folded laser cavity formed by the mirrors M1-7 as seen in Fig. 1. All mirrors are HR coated for 1064 nm. M7 (Plane) is made on an undoped YAG substrate with both sides AR coated for high transmission at 2 – 4 µm. M6 (Plane) acts as the output coupler for the upconverted image. Mirrors M6 and M7 are in a separate compartment to avoid scattered 808 nm pump laser radiation near the non-linear crystal. The path length between M7 and M3 (ROC = 200 mm) is 202 mm and the path length between M1 and M3 is 156 mm. All mirrors except M3 are plane. The 0.27 atm % Nd-doped $YVO_4$ crystal is 3 x 3 x 12 $mm^3$. The 1064 nm laser is pumped by a 4 W BAL at 808 nm. With this system a cw intracavity field of 100 W was realized. The non-linear material is a 5 % Mg:O doped Lithium Niobate crystal produced by Covesion Ltd. on custom order. The crystal contains 5 different poling periods ranging from 21 to 23 µm in steps of 0.5 µm located side by side. Each poling period has a 1x1 mm aperture. The crystal length is 20 mm. The non-linear crystal is positioned in a near circular beam waist of 180 µm at M7. The temperature of the non-linear crystal can be controlled with 0.1 °C accuracy from just above room temperature, to 200°C. It is important to notice that the imaged wavelengths change with the poling period, as well as the crystal temperature. A computer program calculates the required poling period and corresponding crystal temperature to phase-match a desired wavelength. The oven containing the non-linear crystal can be translated. This allows for selection of the required poling period.

The infrared light is passed through an AR-coated Germanium window, to remove radiation below ~2 µm before entering the upconverter unit. Inside the non-linear crystal, the mid-IR light mixes with the laser beam and is upconverted to the near visible region, determined by the phase-match condition for the non-linear process. In order to remove laser light and unwanted wavelengths, the upconverted radiation is passed through several optical filters before entering the camera. Typically these filters include a 1064/532 notch filter with ~98 % transparency at the upconverted wavelengths combined with 750 nm, long pass, and a 850 nm, short pass filter. When imaging wavelengths above 4.2 µm, the filters are exchanged with a long pass 850 nm and a short pass 900 nm. This is necessary, since a given poling period and temperature will simultaneously phase-match wavelengths both above and below 4.2 µm.

**Estimation of quantum efficiency.** The QE is measured using a 1073 K calibrated blackbody cropped by a circular aperture (Ø = 6.7 mm) placed 100 mm from a f = 100 mm IR lens. The power of upconverted light from this source is compared with theoretical predictions[30]. The power of the upconverted light, using a pW power meter, corresponds perfectly with the theoretically predicted power when corrected for transmission losses in the output coupler and filters. The conversion efficiency is 20 % for phase-matched light centered at the optical axis of the laser cavity. The theoretical calculations are based on an effective second order non-linearity of 14 pm/V as specified by the crystal manufacturer (Covesion Ltd.), and a laser beam radius of 180 µm inside the non-linear crystal.

**Noise and single photon measurements.** In order to estimate the noise generated by the non-linear crystal in the upconversion process, we measure the upconverted signal at a specific wavelength (3 µm). The upconverted wavelength selection, is ensured to be 3 µm by use of a narrowband bandpass-filter at 785 nm, with <3 nm bandwidth. An image is acquired at 10 second exposure time at many different crystal temperatures. These images contain 3 different parts with different noise contributions. This is illustrated in Fig. 4b. To subtract the camera-generated noise (particularly clock induced charge), which is much higher than the actual signals, we measure the difference signal from the area within the bandpass-filter and subtract the signal from an area at the same vertical position (which subtracts the same amount of clock induced charge generated by the camera), but well outside the phase-matched area (different horizontal position). For different crystal temperatures the subtracted areas are kept at the same vertical position, but at different horizontal positions following the angular variation of phase-match. The plotted values in Fig. 4a are these average pixel intensity differences. We plot these values for the three different noise levels in the image. This is subsequently compared to the theoretical predicted noise, which is only plotted for the glass reflection part and the central part of the image (i.e. the reflection from the copper heat sink is not plotted).

To measure the single photons in Fig. 4c, we exchanged the Andor Luca S camera with an Andor iXon3 897. Furthermore, we introduced a 4-f image relay system making a phase-conserving image of the non-linear crystal in a position where we can insert an aperture. The aperture is used to make sure that only light coming from the non-linear crystal is allowed to reach the camera. Please note that this aperture and image relaying are not needed during normal operation or low light level imaging – only for optimizing the single photon imaging. After the image relay the image is formed by a lens situated in a 2-f geometry. The entire image relay optics is equivalent to a 2-f imaging with a focal length of 100 mm. From theory, camera pixel size, filter transparency, intracavity field strength etc. we can calculate the expected amount of detected photons per pixel per second to be ~0.5. This corresponds well to the intensity reported by the camera when binning 64 pixels on the camera before readout, and using the calibrated photon readout feature of the iXon3.

**Figures:**

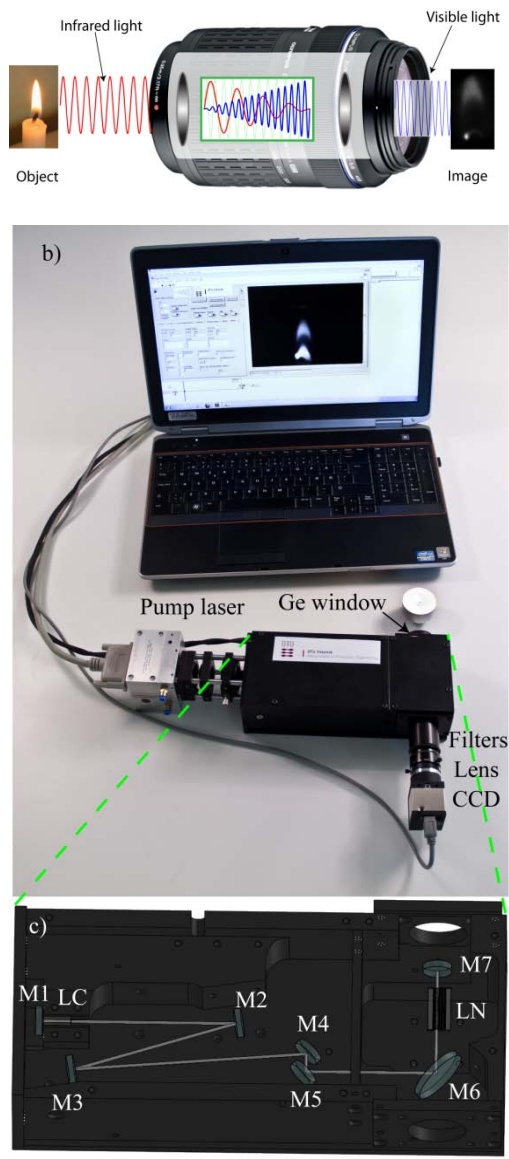

Fig. 1: a) The concept of image upconversion. The mid-IR light is converted in a single pass through a non-linear crystal mixing the mid-IR light with a 1064 nm laser beam. Image information is conserved in the process. b) Photograph of the upconversion device. Mid-IR light, in this case from a candle, is passed through a Germanium window which cuts off visible and NIR light. The infrared light is then mixed with the resonant laser field in a single pass configuration to generate the upconverted light at near visible wavelengths. c) Internal view of image conversion module.

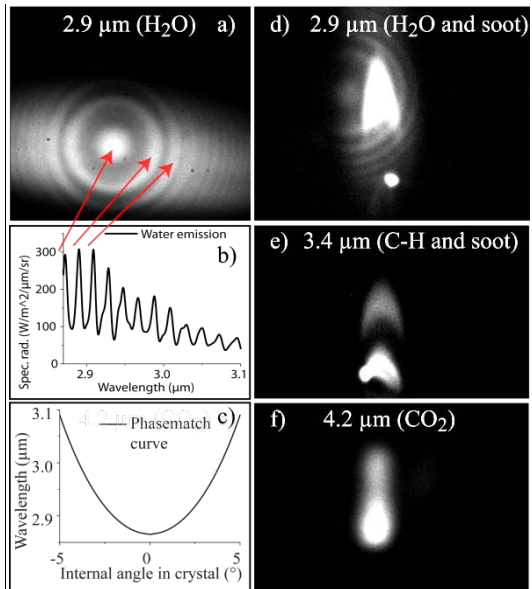

Fig. 2. Spectral imaging. a) An image of a flame from a butane gas burner. The ring pattern in the image matches the spectrum of infrared emission from hot water vapor as shown in image b) and indicated with red arrows. Image c) shows how the phase-matched wavelength in the image changes as function of the internal angle in the crystal. Image d) shows a spectral image of a candle at ~2.9 µm where water vapor is emission is pronounced. The spectral ring pattern as also seen in image a). Blackbody radiation from soot and the wick is also apparent in the image. Image e) shows the blackbody emission from soot and the wick, but also how hot un-combusted hydrocarbons emit light around 3.4 µm. Image f) shows the distribution of $CO_2$ in the candle flame – a hot plume of $CO_2$ can be traced high above the candle (not shown). Note that the individual images are scaled differently.

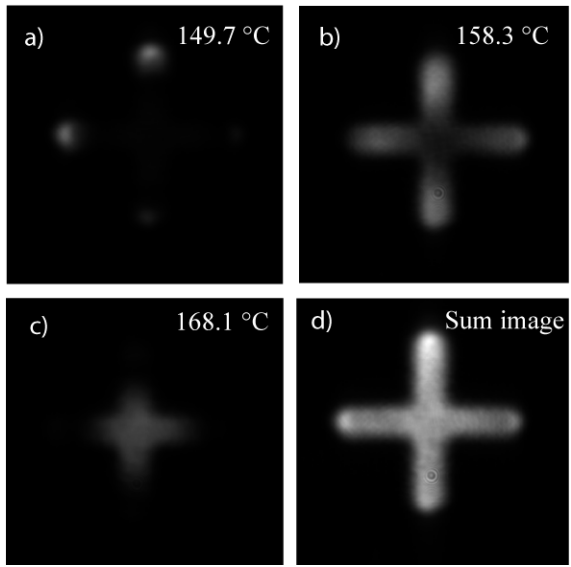

Fig. 3. Monochromatic image reconstruction can be obtained by scanning the phase-match condition combined with post processing of the images. This is demonstrated by adding images, a)-c), for different crystal temperatures, to generate the monochromatic image d). The cross target was illuminated by a mid-IR narrow-band 3 µm laser.

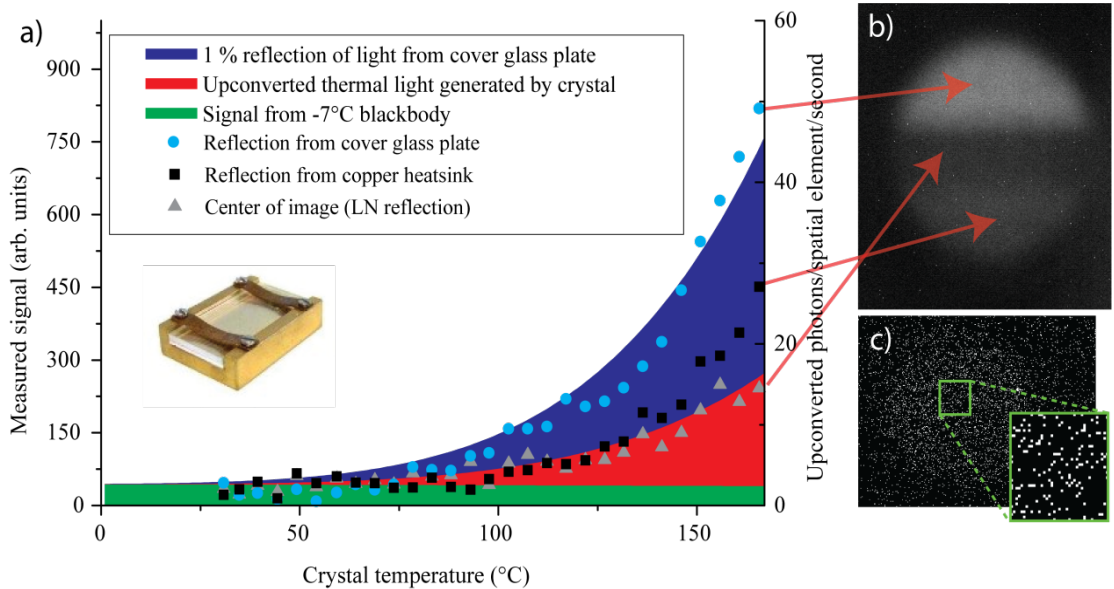

Fig. 4. Noise analysis. A measurement and theoretical calculation of the generated blackbody noise is shown in a). The theoretical calculated signal from a -7°C blackbody is indicated as green. The red area of the curve illustrates how thermal radiation generated by the 0.5 % emissivity of MgO:LN increases as the crystal temperature is increased. The blue area indicates how a 1 % reflection of light emitted from a blackbody source contributes to the signal. In the chosen experimental configuration, this reflection is caused by imperfect AR coating on the end-mirror (M7) in the linear cavity and the thermal radiation is emitted by the glass cover plate above the MgO:LN crystal. Since this reflection originates from the cover plate above the non-linear crystal, it only appears in the topmost part of the image, as seen in b). However, the lower part of the image contains a reflection from the copper heat sink, which has a lower emissivity. The central part of the image contains only the reflection from the transparent crystal, explaining why the noise in this part of the image is much lower. As function of the crystal temperature, we measure the total signal in these three parts of the image, as indicated with circles, squares and triangles. Theory and experiment are seen to correspond well to each other. The virtually noise free performance at room temperature is exploited to measure single

photons as indicated in image c). The white dots indicate (at least) one detected photon and the black parts are pixels with 0 photons. The photon source is a 25°C calibrated blackbody. The imaging area is the circular area, with a denser photon population, and the darker surroundings indicate the noise level caused by imperfect darkness and imperfect filtering of laser photons. A 0.3 second exposure time was used to acquire these single photons at 3 µm with 5 nm bandwidth.